\documentclass[preprintnumbers,article,amsmath,amssymb,floatfix,10pt,prd,twocolumn,
superscriptaddress,nofootinbib]{revtex4-2}
\usepackage{bm}
\usepackage{amsfonts}
\usepackage{latexsym}
\usepackage[utf8]{inputenc}
\usepackage{graphicx}
\usepackage{amsmath}
\usepackage{palatino}
\usepackage{mathpazo}
\usepackage{textcomp}
\usepackage{float}
\usepackage{booktabs}
\usepackage{dcolumn}
\usepackage{ragged2e}
\usepackage{hyperref}
\hypersetup{colorlinks,citecolor=blue}
\hypersetup{colorlinks=true,linkcolor=red,filecolor=magenta,    urlcolor=blue}
\usepackage{amsmath}
\usepackage{xcolor}
\usepackage{orcidlink}
\usepackage{epsfig}
\usepackage{caption}
\usepackage{subcaption}

\usepackage{commath}
\def\jnl@style{\it}
\def\aaref@jnl#1{{\jnl@style#1}}

\def\aaref@jnl#1{{\jnl@style#1}}

\def\aj{\aaref@jnl{AJ}}                   
\def\apj{\aaref@jnl{ApJ}}                 
\def\apjl{\aaref@jnl{ApJ}}                
\def\apjs{\aaref@jnl{ApJS}}               
\def\apss{\aaref@jnl{Ap\&SS}}             
\def\aap{\aaref@jnl{A\&A}}                
\def\aapr{\aaref@jnl{A\&A~Rev.}}          
\def\aaps{\aaref@jnl{A\&AS}}              
\def\mnras{\aaref@jnl{Mon.~Not.~Roy.~Astron.~Soc.}}             
\def\prd{\aaref@jnl{Phys.~Rev.~D}}        
\def\prc{\aaref@jnl{Phys.~Rev.~C}}  
\def\prl{\aaref@jnl{Phys.~Rev.~Lett.}}    
\def\qjras{\aaref@jnl{QJRAS}}             
\def\skytel{\aaref@jnl{S\&T}}             
\def\ssr{\aaref@jnl{Space~Sci.~Rev.}}     
\def\zap{\aaref@jnl{ZAp}}                 
\def\nat{\aaref@jnl{Nature}}              
\def\aplett{\aaref@jnl{Astrophys.~Lett.}} 
\def\apspr{\aaref@jnl{Astrophys.~Space~Phys.~Res.}} 
\def\physrep{\aaref@jnl{Phys.~Rep.}}      
\def\physscr{\aaref@jnl{Phys.~Scr}}       
\def\commat{\aaref@jnl{Comm.~Math.~Phys.}}              
\def\science{\aaref@jnl{Science}}               
\def\cqg{\aaref@jnl{Classical Quant.~Grav.}}            
\def\jpcs{\aaref@jnl{JPCS}}                                     
\def\ijmpd{\aaref@jnl{Int.~J.~Mod.~Phys.~D}}                    
\def\grg{\aaref@jnl{Gen.~Relat.~Gravit.}}               
\def\rpp{\aaref@jnl{Rep.~Prog.~Phys.}}          
\def\npa{\aaref@jnl{Nucl.~Phys.~A}}        
\def\lrr{\aaref@jnl{Living Rev.~Rel.}}                   
\def\jcap{\aaref@jnl{J.~Cosmology Astropart.~Phys.}}    
\def\rmp{\aaref@jnl{Rev.~Mod.~Phys.}}   
\def\epjc{\aaref@jnl{Eur.~Phys.~J.~C}} 
\def\plb{\aaref@jnl{~Phy.~Lett.~B}} 
\def\mpla{\aaref@jnl{Mod.~Phy.~Lett.~A}} 
\def\arxiv{\aaref@jnl{arxiv.org}}

\allowdisplaybreaks[1]

\begin{document}
\color{black}       
\title{Observational constraints on viscous cosmology in $f(T,L_m)$ gravity}

\author{M. Koussour\orcidlink{0000-0002-4188-0572}}
\email[Email: ]{pr.mouhssine@gmail.com}
\affiliation{Department of Physics, University of Hassan II Casablanca, Morocco}

\author{Alnadhief H. A. Alfedeel\orcidlink{0000-0002-8036-268X}}%
\email[Email: ]{aaalnadhief@imamu.edu.sa}
\affiliation{Department of Mathematics and Statistics, Imam Mohammad Ibn Saud Islamic University (IMSIU),
Riyadh 13318, Saudi Arabia}

\author{S. Muminov\orcidlink{0000-0003-2471-4836}}
\email[Email: ]{sokhibjan.muminov@gmail.com}
\affiliation{Mamun University, Bolkhovuz Street 2, Khiva 220900, Uzbekistan}

\author{J. Rayimbaev\orcidlink{0000-0001-9293-1838}}
\email[Email: ]{javlon@astrin.uz}
\affiliation{Institute of Theoretical Physics, National University of Uzbekistan, University Str. 4, Tashkent 100174, Uzbekistan}
\affiliation{Urgench State University, Kh. Alimjan str. 14, Urgench 221100, Uzbekistan}

\begin{abstract}

We investigate the late-time cosmic acceleration within the framework of viscous $f(T,L_m)$ gravity, where the gravitational action depends on both the torsion scalar $T$ and the matter Lagrangian $L_m$. In this context, the Universe is modeled as a bulk viscous fluid, allowing for dissipative effects that generate an effective negative pressure capable of driving acceleration without invoking a cosmological constant. We adopt a simple linear model $f(T,L_m) = \alpha T + \beta L_m$ and assume a constant bulk viscosity coefficient $\zeta = \zeta_0 > 0$. The model parameters are constrained using a joint analysis of recent observational datasets, including 31 Hubble parameter measurements, the Pantheon+ sample of 1701 Type Ia Supernovae, and the latest baryon acoustic oscillation data from DESI, employing a Markov Chain Monte Carlo (MCMC) approach. The best-fit results, $H_0 = 68.16 \pm 0.65$, $\alpha = 1.53^{+0.49}_{-0.61}$, $\beta = 0.40 \pm 0.96$, and $\zeta_0 = 2.15^{+0.69}_{-0.81}$, are consistent with current cosmological observations and indicate that bulk viscosity plays a significant role in the late-time dynamics. The deceleration parameter $q_0 = -0.33 \pm 0.41$ confirms the current accelerated expansion, while the effective equation of state (EoS) evolves from a matter-like regime at high redshift toward a quintessence phase at late times. The $Om(z)$ diagnostic further supports this behavior, suggesting a mild deviation from $\Lambda$CDM toward a dynamical dark energy component. Although information criteria ($\Delta \mathrm{AIC} = 2.2$, $\Delta \mathrm{BIC} = 13.13$) slightly favor the simpler $\Lambda$CDM model, the viscous $f(T,L_m)$ framework remains a viable and physically motivated alternative capable of explaining cosmic acceleration through the combined effects of torsion–matter coupling and viscosity.


\end{abstract}

\maketitle


\section{Introduction}\label{sec1}

Recent advances in observational cosmology have profoundly reshaped our understanding of the Universe’s large-scale dynamics. High-precision measurements from Type Ia Supernovae (SNe Ia) \cite{Perlmutter/1999,Riess/1998,Riess/2004}, the Wilkinson Microwave Anisotropy Probe (WMAP) \cite{Spergel/2003}, baryon acoustic oscillation (BAO) surveys \cite{D.J.,W.J.}, and large-scale structure data \cite{Koivisto/2006,Daniel/2008} have consistently confirmed that the Universe is currently experiencing a phase of accelerated expansion. These observations collectively indicate that the total energy budget of the cosmos is dominated by two unknown components: dark matter (DM) and dark energy (DE), which together account for nearly 95–96\% of the total energy density, while ordinary baryonic matter represents only a small fraction of about 4–5\% \cite{Peebles_2003,Padmanabhan_2003}. Although the standard model of cosmology based on general relativity (GR) successfully explains a wide range of gravitational phenomena at local and solar system scales, it faces significant challenges at galactic and cosmological levels. In particular, GR does not inherently explain the origin or nature of DM and DE, both of which are introduced phenomenologically to reconcile theory with observations. Furthermore, GR predicts the occurrence of spacetime singularities, such as those at the Big Bang or within black holes, where the laws of physics cease to be predictive. These issues suggest that GR, while remarkably robust in the weak-field regime, is not a complete description of gravity and may require modifications or extensions to consistently address the late-time cosmic acceleration and the initial singularity problem.

In recent decades, numerous theoretical frameworks have been developed to explain the accelerating expansion of the Universe and other cosmological anomalies. One of the most successful and straightforward models is the standard $\Lambda$CDM paradigm, where the cosmological constant $\Lambda$ represents DE with a constant equation of state (EoS) parameter $\omega_{\Lambda} = -1$. Although $\Lambda$CDM agrees remarkably well with most cosmological observations, it faces conceptual challenges, notably the fine-tuning and coincidence problems \cite{Weinberg/1989,Zlatev/1999}. Alternatively, it is plausible that Einstein’s GR does not fully describe gravitational interactions on cosmological scales, suggesting that the standard Hilbert–Einstein action should be extended. Within the Riemannian framework, one of the most direct generalizations replaces the Ricci scalar $R$ by an arbitrary function $f(R)$, leading to the well-known $f(R)$ gravity \cite{Buchdahl_1970,Barrow_1983}. These extensions not only broaden the scope of GR but also provide geometric explanations for DM phenomena \cite{Boehmer_2008}. Further developments introduced couplings between geometry and matter, giving rise to theories such as $f(R,L_m)$ gravity \cite{Bertolami_2007,Harko_2008,Harko_2010,Harko_2011}, and later $f(R,\mathcal{T})$ gravity, which includes the trace of the energy–momentum tensor $\mathcal{T}$ as an additional argument \cite{Harko_2011_fRT}. Since GR is inherently a Riemannian theory, another natural direction is to explore alternative geometric foundations of gravity. Extending beyond Riemannian geometry, whose applicability might be limited to local or weak-field regimes, allows the formulation of theories capable of describing cosmic-scale dynamics and the late-time acceleration of the Universe. In this context, a key development was introduced by Weitzenböck, who defined a class of manifolds characterized by vanishing non-metricity $Q = 0$ and zero curvature $R = 0$, while allowing a non-vanishing torsion tensor $T \neq 0$ \cite{Weitzenbock_1923}. When torsion vanishes, the structure reverts to standard Euclidean geometry. The distinctive property of Weitzenböck spaces is the coexistence of zero curvature and non-zero torsion, ensuring distant parallelism (commonly known as teleparallelism), a concept that later inspired Einstein’s attempt to formulate a unified theory of gravitation and electromagnetism \cite{Einstein_1928}.

The teleparallel approach to gravity offers an alternative geometric description of spacetime, in which gravitational interactions are encoded not solely by the metric tensor $g_{\mu\nu}$, but through a set of tetrad fields $e^i_{\mu}$. These tetrads give rise to torsion, which fully accounts for gravitational effects, effectively replacing curvature as the primary geometric quantity. This formulation leads to the teleparallel equivalent of GR (TEGR) \cite{Moller_1961,Pellegrini_1963,Hayashi_1979}, which has been further generalized into $f(T)$ gravity. In this framework, the torsion scalar $T$ plays a role analogous to the Ricci scalar $R$ in GR, while the spacetime remains curvature-free. A key advantage of $f(T)$ theories is that their field equations are second-order, unlike the typically fourth-order equations of $f(R)$ gravity in the metric formulation \cite{Aldrovandi_2013}. Modified teleparallel gravity has seen growing interest in cosmology and astrophysics. Initial studies by Ferraro and Fiorini \cite{Ferraro2007,Ferraro2008} laid the foundation for $f(T)$ models, exploring inflationary scenarios and Born–Infeld-inspired extensions. Subsequent works by Bengochea and Ferraro \cite{Bengochea2009} demonstrated that torsion could drive late-time cosmic acceleration, while Linder \cite{Linder2010} proposed $f(T)$ as an alternative to DE. Extensions of $f(T)$ gravity have been applied to exotic spacetime geometries, such as wormholes \cite{Boehmer2012}, and further generalized to include torsion–matter couplings and $f(T,\mathcal{T})$ constructions \cite{Harko2014b}. Recent developments encompass a wide range of topics, including new classes of viable models \cite{Bahamonde2017}, early-universe constraints like big bang nucleosynthesis \cite{Capozziello2017}, the equivalence to GR \cite{Combi2018}, gravitational wave predictions \cite{Farrugia2018}, inflationary dynamics \cite{Awad2018}, cosmic acceleration with viscous fluids \cite{Myrzakulov2025}, modeling of compact stars \cite{Khokhar2025}, and Hamiltonian analyses \cite{Bajardi2025}. Recently, Harko et al. \cite{Harko2014a} proposed extensions of teleparallel gravity in which the matter Lagrangian couples non-minimally to the torsion scalar, leading to modified gravitational dynamics through direct torsion–matter interactions. In particular, models of the form $(1+\lambda f(T))L_m$ introduce a multiplicative coupling between geometry and matter within the $f(T)$ framework. In contrast, in the present work, we consider a theory belonging to the class of $f(T,L_m)$ gravity, where the gravitational Lagrangian depends explicitly on both the torsion scalar $T$ and the matter Lagrangian $L_m$. Even for the linear choice, this framework is structurally different from non-minimal torsion-matter coupling models and leads to distinct field equations and cosmological dynamics. The linear model adopted here should therefore be regarded as a minimal realization of $f(T,L_m)$ gravity rather than a special case of previously studied nonminimal coupling theories.

In the majority of standard and modified gravity studies, the cosmic fluid is commonly modeled as perfect, neglecting any viscous effects. From the standpoint of fluid dynamics, however, this assumption is somewhat idealized, as viscosity naturally arises in many physical systems, even in homogeneous and boundary-free conditions. The deviations from thermal equilibrium introduce two primary viscosity coefficients: shear viscosity $\eta$ and bulk viscosity $\zeta$, typically considered at first-order perturbations in the non-causal framework developed by Eckart \cite{Eckart/1940}. To ensure consistency with causality and special relativity, one must incorporate second-order corrections, as formalized in the pioneering works of Müller \cite{Muller/1967}, Israel \cite{Israel/1976}, and Israel \& Stewart \cite{Israel/1979}. In cosmological settings, the assumption of spatial isotropy often allows shear viscosity to be neglected, making bulk viscosity the dominant effect in the fluid dynamics. Recent investigations by Brevik and collaborators \cite{Brevik/2005,Brevik/2006a,Brevik/2006b,Brevik/2012} have explored the role of bulk viscosity within modified gravity frameworks, highlighting its potential influence on the Universe’s late-time accelerated expansion. In this work, we aim to explore the cosmological implications of a viscous fluid within the framework of $f(T,L_m)$ gravity. Specifically, we consider a linear model of the form $f(T,L_m) = \alpha T + \beta L_m$, incorporating a constant bulk viscosity coefficient $\zeta = \zeta_0 > 0$. Our primary objective is to investigate whether the combined effects of torsion–matter coupling and bulk viscosity can account for the observed late-time acceleration of the Universe without invoking a cosmological constant. To this end, we constrain the model parameters using recent observational datasets, including Hubble parameter measurements, the Pantheon+ SNe Ia sample, and DESI BAO data, and examine the resulting cosmological dynamics through reconstructed parameters such as $q(z)$, $\omega_v(z)$, and $Om(z)$.

The paper is organized as follows. In Sec. \ref{sec2}, we present the basic formalism of $f(T,L_m)$ gravity and derive the corresponding field equations. Sec. \ref{sec3} is dedicated to the study of viscous fluid dynamics in a spatially flat FLRW background, where we obtain the modified Friedmann equations and discuss the cosmological solutions. In Sec. \ref{sec4}, we describe the observational datasets used to constrain the model parameters and perform a comparative statistical analysis with the standard $\Lambda$CDM model. Sec. \ref{sec5} focuses on the cosmological evolution and related observables. Finally, in Sec. \ref{sec6}, we summarize the main findings and present our concluding remarks.

\section{Basic formalism of $f(T,L_m)$ gravity}\label{sec2}

In this paper, we consider a 4-dimensional differentiable manifold endowed with a tetrad (vierbein) field $e^{A}{}_{\mu}(x)$, which forms an orthonormal basis for the tangent space at each spacetime point. Greek indices $(\mu,\nu,\dots)$ denote spacetime coordinates, while Latin indices $(A,B,\dots)$ refer to the tangent space, equipped with the Minkowski metric $\eta_{AB}=\mathrm{diag}(+1,-1,-1,-1)$. The spacetime metric is reconstructed from the tetrad components as
\begin{equation}
    g_{\mu\nu}=\eta_{AB}\,e^{A}{}_{\mu}e^{B}{}_{\nu},\qquad e\equiv\det\big(e^{A}{}_{\mu}\big)=\sqrt{-g}.
\end{equation}

In the teleparallel description of gravity, torsion replaces curvature as the fundamental geometrical quantity. The relevant connection is the curvature-free Weitzenböck connection \cite{Weitzenbock_1923}, given by
\begin{equation}
\Gamma^{\lambda}{}_{\mu\nu}=e_{A}{}^{\lambda}\partial_{\nu}e^{A}{}_{\mu}    
\end{equation}

Its associated torsion tensor, encoding the gravitational degrees of freedom, reads
\begin{equation}
    T^{\rho}{}_{\mu\nu}\equiv\Gamma^{\rho}{}_{\nu\mu}-\Gamma^{\rho}{}_{\mu\nu}
    =e_{A}{}^{\rho}\big(\partial_{\mu}e^{A}{}_{\nu}-\partial_{\nu}e^{A}{}_{\mu}\big).
\end{equation}

The difference between the Weitzenböck and Levi–Civita connections is measured by the contorsion tensor,
\begin{equation}
    K^{\mu\nu}_{\phantom{\mu\nu}\rho}\equiv -\frac{1}{2}\big(T^{\mu\nu}_{\phantom{\mu\nu}\rho}-T^{\nu\mu}_{\phantom{\mu\nu}\rho}-T_{\rho}^{\mu\nu}\big),
\end{equation}
and a related auxiliary tensor, the superpotential $S_{\rho}{}^{\mu\nu}$, is defined as 
\begin{equation}
    S_{\rho}{}^{\mu\nu}\equiv\frac{1}{2}\Big(K^{\mu\nu}_{\phantom{\mu\nu}\rho}+\delta^{\mu}_{\rho}T^{\alpha\nu}_{\phantom{\alpha\nu}\alpha}-\delta^{\nu}_{\rho}T^{\alpha\mu}_{\phantom{\alpha\mu}\alpha}\Big).
\end{equation}

The torsion scalar, the teleparallel analogue of the Ricci scalar, is constructed from these tensors as
\begin{eqnarray}
    T &\equiv& T^{\rho}{}_{\mu\nu} S_{\rho}{}^{\mu\nu}
    \nonumber \\
      &=& \frac{1}{4} T^{\rho\mu\nu} T_{\rho\mu\nu}
          + \frac{1}{2} T^{\rho\mu\nu} T_{\nu\mu\rho}
          - T_{\rho\mu}{}^{\rho} T^{\nu\mu}{}_{\nu} \,.
    \label{TorsionScalar}
\end{eqnarray}

Further, the matter Lagrangian depends on both the matter fields and the tetrad, and its variation defines the energy–momentum tensor in the tetrad formalism \cite{Aldrovandi_2013},
\begin{equation}\label{3}
    \mathcal{T}_{A}{}^{\mu}\equiv -\frac{1}{e}\frac{\delta(e L_{m})}{\delta e^{A}{}_{\mu}}.
\end{equation}

To incorporate a non-minimal coupling between torsion and matter, we consider a gravitational action where the Lagrangian density depends on both the torsion scalar $T$ and the matter Lagrangian $L_m$,
\begin{equation}\label{action}
    S=\frac{1}{2\kappa^{2}}\int d^{4}x\,e\,f(T,L_{m})
    +\int d^{4}x\,e\,L_{m}.
\end{equation}
where $\kappa^{2}=8\pi G$. By varying the action with respect to the tetrad components $e^{A}{}_{\mu}$, one obtains the gravitational field equations that describe the interplay between matter, energy, and the geometrical structure of spacetime,
\begin{eqnarray} \label{field}
&&\Big[ e^{-1} \partial_\mu \big( e\, e_A{}^{\rho} S_{\rho}{}^{\mu\nu} \big)
      - e_A{}^{\lambda} T^{\rho}{}_{\mu\lambda} S_{\rho}{}^{\nu\mu} \Big] f_{T} + e_A{}^{\rho} S_{\rho}{}^{\mu\nu} \nonumber \\
      &&\quad \times
      \left( f_{TT}\,\partial_\mu T + f_{TL}\,\partial_\mu L_{m} \right)
      - \frac{1}{4} f_{L}\, e_A{}^{\rho}
      \left( \mathcal{T}_{\rho}{}^{\nu} + L_{m} \right)  \nonumber \\
&&\quad + e_A{}^{\nu} \left( \frac{f}{4} \right)
      = \frac{\kappa^{2}}{2}\, e_A{}^{\rho} \mathcal{T}_{\rho}{}^{\nu} \, .
\end{eqnarray}

Here,  $f_{T}\equiv\frac{\partial f}{\partial T}$, $f_{L}\equiv\frac{\partial f}{\partial L_{m}}$, $f_{TT}\equiv\frac{\partial^{2} f}{\partial T^{2}}$, and $f_{TL} \equiv \frac{\partial^{2} f}{\partial T \, \partial L_{m}}$. Eq. (\ref{field}) generalizes the field equations of teleparallel gravity by introducing explicit coupling between the torsional and matter sectors through the $f_L$ term, and higher-order corrections via $f_{TT}$ and $f_{TL}$ when the Lagrangian is nonlinear in $T$ or $L_m$.

\section{Viscous fluid dynamics} \label{sec3}

To explore the cosmological behavior within the framework of $f(T, L_m)$ gravity, we adopt a spatially flat, homogeneous, and isotropic spacetime described by the Friedmann–Lemaître–Robertson–Walker (FLRW) metric \cite{ryden/2003},
\begin{equation}
ds^2 = dt^2 - a(t)^2 \left( dx^2 + dy^2 + dz^2 \right),
\end{equation}
where $a(t)$ is the cosmic scale factor. The corresponding diagonal tetrad field is chosen as $e_{\mu}^A = \mathrm{diag}(1, a(t), a(t), a(t))$, leading to the torsion scalar,
\begin{equation}
T = -6 H^2,    
\end{equation}
where $H \equiv \frac{\dot{a}}{a}$ denotes the Hubble parameter, which quantifies the rate of cosmic expansion. $H \equiv \frac{\dot{a}}{a}$.

Moreover, we consider the Universe to be filled with a bulk viscous fluid, which deviates from the ideal perfect-fluid assumption due to the presence of internal frictional effects that arise during cosmic expansion. In this framework, the cosmic medium is modeled as an imperfect fluid, whose energy–momentum tensor takes the general form
\begin{equation}
\mathcal{T}^{\mu}{}_{\nu} = \mathrm{diag}(\rho, -p_v, -p_v, -p_v),
\end{equation}
where $\rho$ represents the energy density and $p_v$ denotes the effective pressure that includes the bulk viscous contribution. The presence of bulk viscosity modifies the equilibrium pressure $p$ as $p_v = p - 3\zeta H$, where $\zeta$ is the bulk viscosity coefficient \cite{Sasidharan/2015}. The term $3\zeta H$ quantifies the internal dissipative effect caused by the expansion of the Universe, effectively introducing a negative pressure component that can drive accelerated expansion. In the case of pressureless (dust-like) matter, where ($p = 0$), the effective pressure simplifies to $p_v = -3\zeta H$. In general, the bulk viscosity coefficient $\zeta$ may be expressed as a function of the cosmic expansion rate and its time variation, incorporating possible dependencies on both the Hubble parameter and its derivative. This is often represented as \cite{Ren/2006}
\begin{equation}
\zeta = \zeta_0 + \zeta_1 H + \zeta_2\left(\frac{\dot{H}}{H} + H\right),
\end{equation}
where $\zeta_0$, $\zeta_1$, and $\zeta_2$ are phenomenological constants characterizing different contributions to the viscous behavior of the cosmic fluid. For simplicity and to allow analytical progress in the cosmological analysis, we consider the minimal case in which the viscosity remains constant throughout cosmic evolution, i.e., $\zeta = \zeta_0 > 0$, while setting $\zeta_1 = \zeta_2 = 0$. 

Substituting the FLRW metric and tetrad into the general field equations \eqref{field} yields the modified Friedmann equations for $f(T, L_m)$ gravity,
\begin{align}
\label{F1}
3 H^{2} &= \frac{1}{4 f_{T}} \Big[ 16 \pi G\rho - f + f_{L} (\rho + L_{m}) \Big], \\
2 \dot{H} &=  \dfrac{-8\pi G (\rho + p_v) - 2 H f_{TL} \dot{L}_m - f_L (p_v + L_m)}{f_T - 12 H^2 f_{TT}},
\label{F2}
\end{align}
where the overdot denotes differentiation with respect to cosmic time $t$.

To derive exact cosmological solutions, the functional form of $f(T, L_m)$ must be specified. Since the gravitational dynamics depend on both the torsion scalar $T$ and the matter Lagrangian $L_m$, distinct forms of $f(T, L_m)$ lead to different cosmological evolutions. In this work, we focus on a simple but physically relevant linear model,
\begin{equation}
f(T,L_m) = \alpha T + \beta L_m,
\end{equation}
where $\alpha$ and $\beta$ are constant parameters. Here, $\alpha$ characterizes the gravitational sector by scaling the torsion contribution, while $\beta$ determines the magnitude of the coupling between torsion and matter. For $\alpha = 1$ and $\beta = 0$, the theory reduces to standard GR with minimal matter coupling. Therefore, the modified Friedmann equations, by taking $L_m = -\rho$ for a viscous cosmic fluid, reduce to 
\begin{align}
\label{model_F1}
6 \alpha H^2 &= (\beta + 16 \pi) \rho, \\
2 \alpha \dot{H} &= (\beta - 8\pi)\rho + 3(\beta + 8\pi)\zeta_0 H.
\label{model_F2}
\end{align}

By combining the two modified Friedmann equations (\ref{model_F1}) and (\ref{model_F2}), we derive the following first-order differential equation for $H$ as
\begin{equation} \label{dHt}
\dot{H} = A H^2 + B H,
\end{equation}
where
\begin{equation}
A \equiv 3\,\frac{\beta - 8\pi}{\beta + 16\pi}, \qquad
B \equiv \frac{3 (\beta + 8\pi) \zeta_0}{2\alpha}. 
\end{equation}

In cosmology, redshift $z$ relates to the scale factor via $1+z = a_0/a(t)$, with $a_0=1$. Time derivatives convert as $d/dt = -(1+z) H(z) d/dz$, so $\dot H = -(1+z) H dH/dz$. Substituting this into Eq. (\ref{dHt}) gives the Hubble evolution in terms of redshift,
\begin{equation}
\frac{dH}{dz} + \frac{A}{1+z}H = -\frac{B}{1+z}.   
\end{equation}

By integrating with respect to $z$, the solution becomes
\begin{equation} \label{Hz}
H(z)= -\frac{B}{A} + \Big(H_0+\frac{B}{A}\Big)(1+z)^{-A}.
\end{equation}

Here, $H_0\equiv H(z=0)$ denotes the current value of the Hubble parameter. The solution describes the Hubble parameter in a viscous $f(T,L_m)$ cosmology, where $A$ controls the redshift scaling and $B$ encodes the bulk viscosity effect. While modified gravity theories and bulk viscosity have independently been proposed to explain the late-time accelerated expansion of the universe, it is physically well motivated to investigate scenarios in which both mechanisms coexist. In the framework of $f(T,L_m)$ gravity with the linear choice, the non-viscous limit ($\zeta_0=0$) leads to a simple power-law evolution of the Hubble parameter, $H(z)=H_0(1+z)^{-A}$, corresponding to a single expansion phase that is either always decelerated or always accelerated, depending on the value of the coupling parameter $\beta$. Such a behavior cannot account for the observed transition from a matter-dominated decelerated era to the current accelerated phase. The inclusion of bulk viscosity ($\zeta_0\neq0$) qualitatively alters the cosmological dynamics. In this case, the Hubble parameter acquires an additional constant contribution, $-B/A$, which encodes the viscous effect and acts as an effective dark energy term at late times. Thus, the deceleration parameter naturally evolves from positive values at high redshift to negative values at low redshift, allowing for a smooth transition between decelerated and accelerated expansion within a single unified framework. At low redshift ($z\to0$), the solution reproduces the present value of the Hubble parameter, $H_0\equiv H(z=0)$, while at high redshift ($z\gg1$) the expansion is governed by the redshift-dependent term, recovering the standard matter-dominated scaling for $\alpha=1$ and $\beta=0$. Therefore, viscosity is not a redundant ingredient but a necessary component for reproducing a realistic cosmic history in $f(T,L_m)$ gravity, simultaneously capturing early-time deceleration and late-time acceleration.

\section{Observational constraints} \label{sec4}

To assess the observational viability of the proposed viscous $f(T, L_m)$ cosmological framework, we perform a comprehensive statistical analysis using a combination of recent datasets. Specifically, we employ 31 Hubble parameter measurements obtained from the cosmic chronometer (CC) technique, the Pantheon+ compilation of 1701 SNe Ia covering a wide redshift range, and the latest BAO constraints from the DESI survey. In addition, the parameter estimation is carried out through a Markov Chain Monte Carlo (MCMC) approach using the \texttt{emcee} sampler \cite{emcee}, with 500 walkers evolved over 5000 iterations and the first 200 steps discarded as burn-in to ensure convergence. Further, we adopt uniform (flat) priors on all model parameters to ensure unbiased exploration of the parameter space. Specifically, the present Hubble parameter is varied within $H_0 \in [60, 80]$ km/s/Mpc. The model parameters are assigned the following ranges: $\alpha \in [-10, 10]$, $\beta \in [-10, 10]$, and the bulk viscosity coefficient $\zeta_0 \in [0, 10]$.

\subsection{$H(z)$, Pantheon+, and DESI datasets}

The $H(z)$ dataset, derived from the differential age method applied to passively evolving galaxies \cite{Moresco/2012,Moresco/2015,Moresco/2016}, provides a direct probe of the expansion rate via the relation $H(z) = -\frac{1}{1+z}\frac{dz}{dt} $. The associated likelihood is expressed as $\chi^2_{H(z)} = \Delta H^T C_{H(z)}^{-1} \Delta H$, where $\Delta H = H_{\rm obs, i} - H_{\rm th}(z)$, and $C_{H(z)}$ includes both statistical and systematic uncertainties \cite{Moresco/2020}.

The Pantheon+ (PP) sample \cite{Scolnic_2022}, which extends over the redshift range $0.001 < z < 2.26$, constrains the luminosity distance $d_L(z) = c(1+z)\int_0^z \frac{dz'}{H(z')}$. The theoretical distance modulus is given by $\mu_{\rm th} = 5 \log_{10}[d_L(z)/10{\rm pc}]$, and its chi-square function is $\chi^2_{\rm SNe} = \Delta \mu^T (C_{\rm stat+syst}^{\rm SNe})^{-1} \Delta \mu$ \cite{Brout_2022}. Here, $\Delta \mu = \mu_{\rm obs, i} - \mu_{\rm th}(z)$, and $C_{\rm stat+syst}^{\rm SNe}$ denotes the complete covariance matrix accounting for both statistical and systematic uncertainties of the SNe Ia sample.

Finally, the first-year DESI BAO results \cite{BAO1,BAO2,BAO3} are included to further constrain the cosmic expansion history. These measurements probe the transverse and radial distance indicators $d_M = D_L/(1+z)$ and $d_H = c/H(z)$, as well as the volume-averaged distance $d_V = [z d_M^2 d_H]^{1/3}$, all normalized to the sound horizon at the drag epoch $r_d$, computed via the Eisenstein–Hu formula \cite{D.J.}. The corresponding chi-square is $\chi^2_{\rm DESI} = \sum_i [(X_i^{\rm th} - X_i^{\rm obs}) / \sigma_{X_i}]^2$, where $X = {d_M/r_d, d_H/r_d, d_V/r_d}$.

The total likelihood function for the combined analysis is obtained by assuming the datasets are statistically independent, so that the joint likelihood is given by the product of the individual likelihoods:
\begin{equation}
\mathcal{L}_{\text{tot}} = \mathcal{L}_{H(z)} \times \mathcal{L}_{\text{SNe}} \times \mathcal{L}_{\text{DESI}}.
\end{equation}

Equivalently, in terms of the chi-square statistic, this corresponds to the sum of the individual chi-square contributions,
\begin{equation}
-2\ln \mathcal{L}_{\rm tot} = \chi^2_{\text{tot}} = \chi^2_{H(z)} + \chi^2_{\text{SNe}} + \chi^2_{\text{DESI}}.
\end{equation}

\begin{table}[h]
\centering
\begin{tabular}{lcc}
\hline\hline
Parameter & $f(T,L_m)$ model & $\Lambda$CDM \\
\hline
$H_0$ [km s$^{-1}$ Mpc$^{-1}$] & $68.16 \pm 0.65$ & $68.62 \pm 0.54$ \\
$\alpha$ & $1.53^{+0.49}_{-0.61}$ & --- \\
$\beta$  & $0.40\pm 0.96$ & --- \\
$\zeta_0$ & $2.15^{+0.69}_{-0.81}$ & --- \\
$\Omega_{m}^0$ & --- & $0.301 \pm 0.010$ \\
$q_0$ & $-0.33 \pm 0.41$ & $-0.549 \pm 0.015$ \\
$z_t$ & $0.88^{+2.10}_{-0.78}$ & $0.67 \pm 0.03$ \\
$\omega_0$ & $-0.52 \pm 0.26$ & $-0.70 \pm 0.01$ \\ 
$\chi^2_{\min}$ & $1940.05$ & $1941.85$ \\
$k$ & $4$ & $2$ \\
$N_{tot}$ & $1744$ & $1744$ \\
$\rm AIC$ & $1948.05$ & $1945.85$ \\
$\Delta \rm AIC$ & $2.2$ & $0$ \\
$\rm BIC$ & $1969.91$ & $1956.78$ \\
$\Delta \rm BIC$ & $13.13$ & $0$ \\
\hline
\end{tabular}
\caption{The table lists the best-fit values and corresponding uncertainties of the free parameters obtained from the joint $H(z)$+SNe+DESI analysis. The values of $\Delta \mathrm{AIC}$ and $\Delta \mathrm{BIC}$ quantify the relative statistical performance of the two models.}
\label{tab}
\end{table}

\begin{widetext}

\begin{figure}[h]
    \centering
    \includegraphics[width=0.9\linewidth]{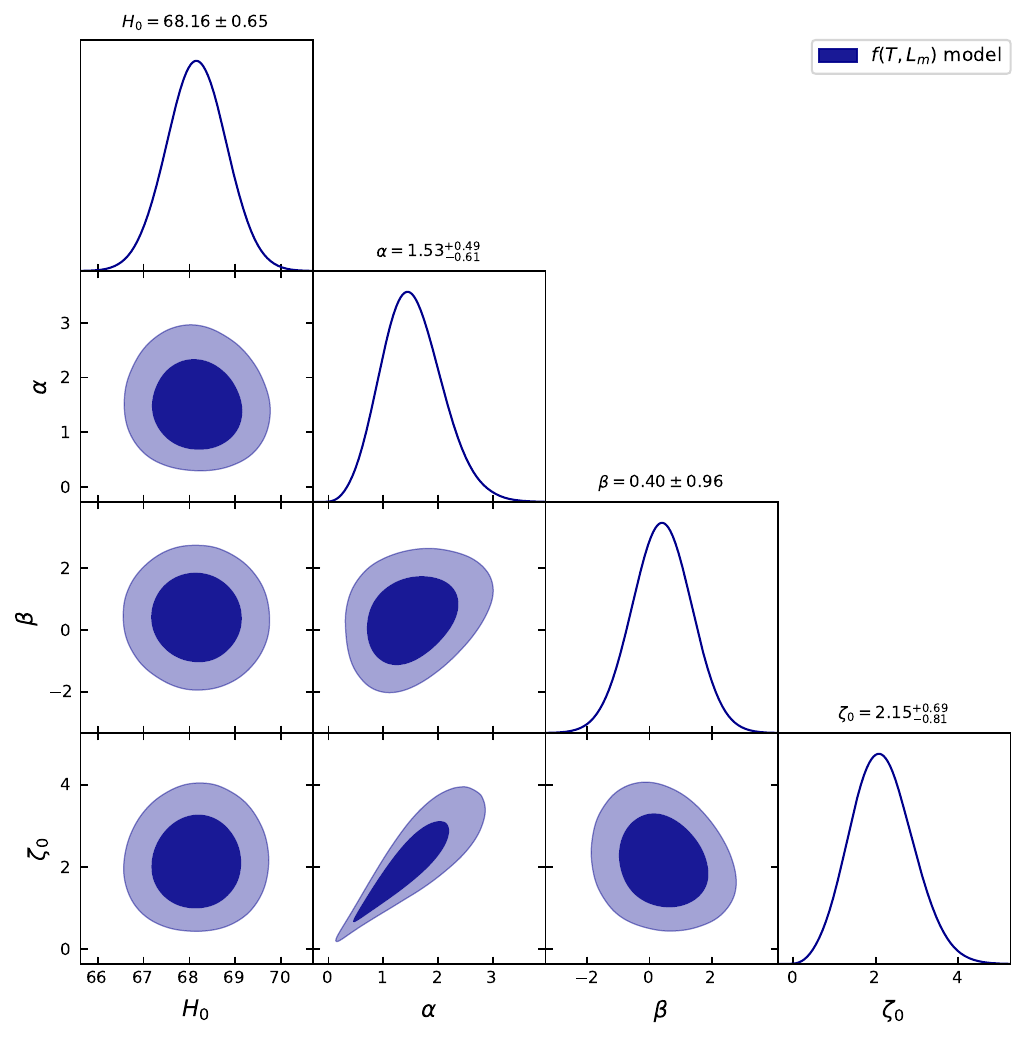}
    \caption{Joint likelihood contours for the viscous $f(T, L_m)$ and $\Lambda$CDM models using the combined $H(z)$+SNe+DESI datasets. The contours correspond to the $1\sigma$ and $2\sigma$ confidence regions for the free parameters.}
    \label{F_fTLm}
\end{figure}

\begin{figure}[h]
    \centering
    \includegraphics[width=0.9\linewidth]{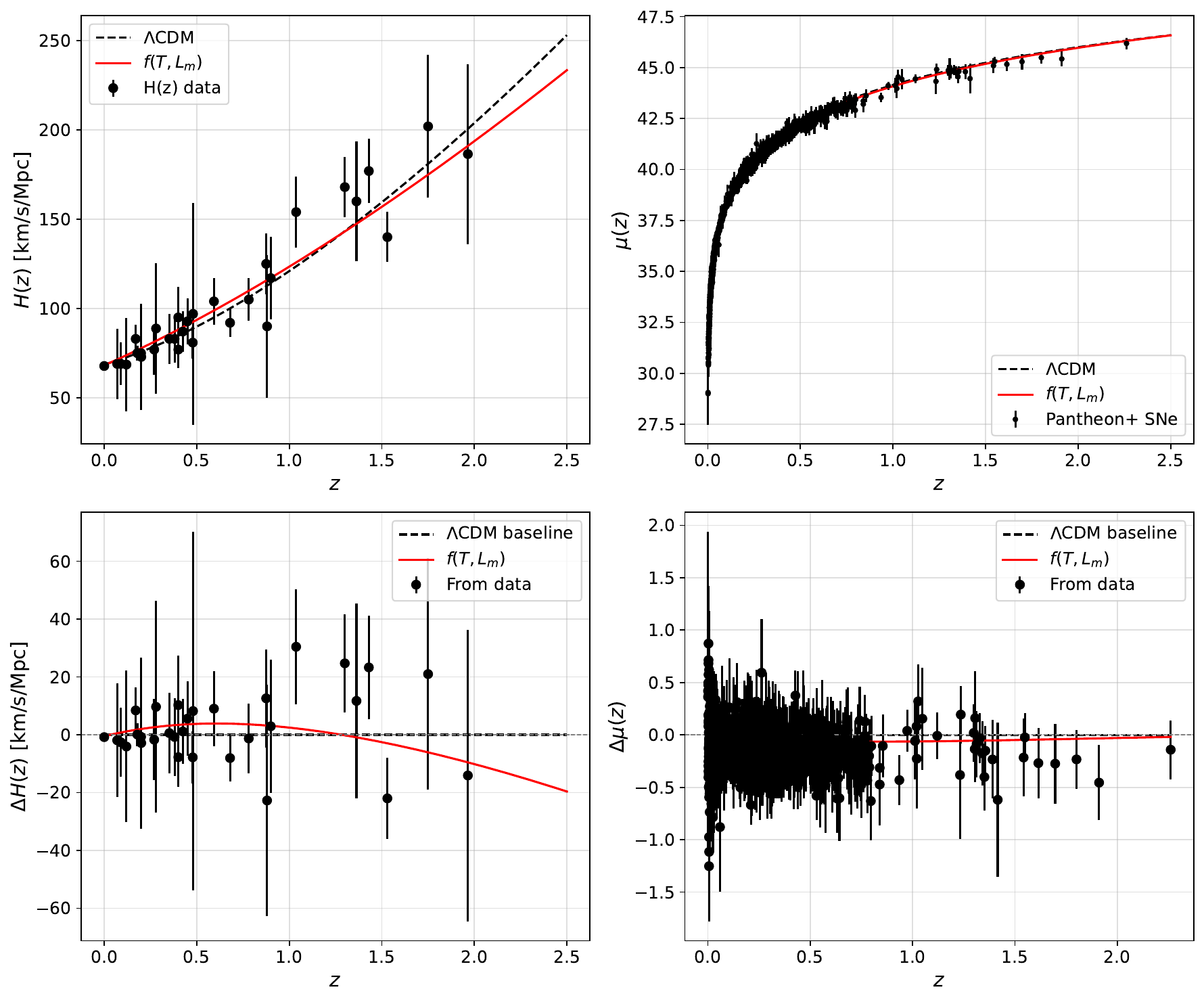}
    \caption{Comparison of the viscous $f(T,L_m)$ model with $\Lambda$CDM: (top-left) Hubble parameter $H(z)$, (top-right) distance modulus $\mu(z)$, (bottom-left) residual differences $\Delta H(z)$, and (bottom-right) residual differences $\Delta \mu(z)$ as functions of redshift.}
    \label{F_comp}
\end{figure}

\end{widetext}

\subsection{MCMC estimation of $H_0$, $\alpha$, $\beta$, and $\zeta_0$}

We perform a joint constraint on the four free parameters $\{H_0, \alpha, \beta, \zeta_0\}$, with the resulting best-fit values summarized in Tab. \ref{tab}. The associated posterior distributions, including the 1$\sigma$ and 2$\sigma$ confidence intervals, are displayed in Fig. \ref{F_fTLm}. Based on the joint analysis of $H(z)$, Pantheon+ SNe Ia, and DESI BAO datasets, the viscous $f(T,L_m)$ model yields $H_0 = 68.16 \pm 0.65$ km/s/Mpc, consistent with Planck CMB measurements \cite{Planck_2015,Planck_2018} and slightly lower than local SH0ES estimates \cite{Riess_2016,Riess_2022}. The gravitational parameter $\alpha = 1.53^{+0.49}_{-0.61}$ governs the effective torsion contribution in the theory. The value $\alpha > 1$ suggests an enhanced torsional effect relative to standard teleparallel gravity $(\alpha = 1)$, which could affect the expansion dynamics at late times. The matter-torsion coupling parameter $\beta = 0.40 \pm 0.96$ remains weakly constrained, with its central value compatible with minimal coupling. The positive bulk viscosity coefficient $\zeta_0 = 2.15^{+0.69}_{-0.81}$ supports the presence of bulk viscous effects, which act as an effective pressure opposing the cosmic contraction and contributing to late-time acceleration. The results indicate that the viscous $f(T,L_m)$ model is compatible with current cosmological observations. The bulk viscosity and torsional modifications can provide additional acceleration without conflicting with $H_0$ measurements. The constraints on $\beta$ and $\alpha$ suggest that deviations from standard GR are mild, while viscosity plays a more pronounced role in driving cosmic acceleration.

\subsection{Observational comparison with $\Lambda$CDM}

To perform a consistent comparison between the viscous $f(T, L_m)$ model and the standard $\Lambda$CDM framework, we define the residual quantities $\Delta H$ and $\Delta \mu$, which quantify the deviation of the model predictions from those of $\Lambda$CDM. Specifically, the Hubble residual is expressed as
\begin{equation}
\Delta H(z) = H(z) - H_{\Lambda \mathrm{CDM}}(z),
\end{equation}
where $H_{\Lambda \mathrm{CDM}}(z) = H_0 \sqrt{\Omega_{m}^0 (1+z)^3 + (1-\Omega_{m}^0)}$ ($\Omega_{m}^0$ is the matter density parameter) corresponds to the Hubble parameter in the $\Lambda$CDM model. Similarly, the SNe Ia distance modulus residual is defined by
\begin{equation}
\Delta \mu(z) = \mu(z) - \mu_{\Lambda \mathrm{CDM}}(z),
\end{equation}
where $\mu_{\Lambda \mathrm{CDM}}(z)$ denotes the theoretical distance modulus in the $\Lambda$CDM scenario. In Fig. \ref{F_comp}, we show the performance of the viscous $f(T,L_m)$ model against observational data using the best-fit parameters from the joint $H(z)$+SNe+DESI analysis. The top-left panel shows the Hubble parameter $H(z)$ as a function of redshift, where the model (red curve) closely follows the observed $H(z)$ data points and slightly deviates from the $\Lambda$CDM prediction (dashed line) at intermediate redshifts, reflecting the effect of viscosity and torsion-matter coupling on the expansion rate. The top-right panel displays the distance modulus $\mu(z)$ for the Pantheon+ SNe Ia dataset, showing excellent agreement between the model and observations across the full redshift range, with negligible deviations from $\Lambda$CDM. The bottom-left panel presents the residuals $\Delta H(z)$, indicating that the deviations of the model from data are mostly within the 1$\sigma$ uncertainties, though the viscous term slightly modifies the high-redshift behavior. Finally, the bottom-right panel shows the residuals $\Delta \mu(z)$ for SNe Ia, confirming that the model reproduces the luminosity distance with high accuracy, and the residuals remain centered around zero. Finally, these figures demonstrate that the viscous $f(T,L_m)$ model with the best-fit parameters provides a reliable description of the observed expansion history, capturing both $H(z)$ and SNe Ia measurements while remaining consistent with $\Lambda$CDM within observational uncertainties.

\subsection{Statistical comparison with $\Lambda$CDM (AIC and BIC)}

To examine the statistical consistency of the viscous $f(T, L_m)$ model relative to the standard $\Lambda$CDM cosmology, we employ two widely used model selection tools: the Akaike Information Criterion (AIC) \cite{AIC1} and the Bayesian Information Criterion (BIC) \cite{AIC2}. Both criteria combine the goodness of fit and model parsimony, penalizing additional free parameters that do not significantly improve the fit. The AIC is defined as \cite{AIC3,AIC4,AIC5}
\begin{equation}
\mathrm{AIC} = \chi^2_{\min} + 2k,
\end{equation}
where $\chi^2_{\min}$ is the minimum chi-square from the observational fit and $k$ represents the number of model parameters. Similarly, the BIC takes the form \cite{AIC4,AIC5,AIC6}
\begin{equation}
\mathrm{BIC} = \chi^2_{\min} + k \ln N_{tot}.
\end{equation}
where $N_{\mathrm{tot}}$ denotes the total number of observational data points.

Smaller AIC and BIC values signify a better compromise between model accuracy and simplicity. The relative performance between the viscous $f(T, L_m)$ and $\Lambda$CDM models are quantified through the differences,
\begin{align}
\Delta \mathrm{AIC} &= \mathrm{AIC}_{f(T,L_m)} - \mathrm{AIC}_{\Lambda \mathrm{CDM}}, \\
\Delta \mathrm{BIC} &= \mathrm{BIC}_{f(T,L_m)} - \mathrm{BIC}_{\Lambda \mathrm{CDM}}.
\end{align}

Following the standard interpretative scale \cite{AIC7}, values of $\Delta < 2$ indicate statistical consistency between models, $2 \leq \Delta < 6$ imply moderate evidence against the one with larger AIC/BIC, and $\Delta > 10$ provides strong evidence against it. These indicators are evaluated for the joint datasets, including 31 cosmic chronometer Hubble data points, 1701 Pantheon+ SNe Ia, and DESI BAO measurements, with the results summarized in Tab. \ref{tab}. We observe that the viscous $f(T,L_m)$ model yields $\Delta \mathrm{AIC} = 2.2$ and $\Delta \mathrm{BIC} = 13.13$ relative to the $\Lambda$CDM cosmology. In this case, indicating that the viscous $f(T,L_m)$ framework provides a fit comparable to $\Lambda$CDM, though slightly less favored when model complexity is considered. However, the larger $\Delta \mathrm{BIC}$ value of 13.13 exceeds the threshold of 10, which represents strong evidence against the viscous $f(T,L_m)$ model in the Bayesian sense. This distinction arises because the BIC penalizes additional parameters more severely than the AIC. Finally, while both models produce nearly equivalent goodness-of-fit statistics ($\chi^2_{\min}$), the information criteria collectively indicate that the simplicity of the $\Lambda$CDM model is statistically preferred over the more parameterized viscous $f(T,L_m)$ model.

\section{Cosmological evolution and observables} \label{sec5}

Now, we investigate the cosmological behavior of the viscous $f(T,L_m)$ model by using the best-fit parameters obtained from the combined $H(z)$, Pantheon+ SNe Ia, and DESI datasets. In particular, we analyze the redshift evolution of key dynamical quantities such as the deceleration parameter $q(z)$, the effective equation of state (EoS) parameter $\omega(z)$, and the $Om(z)$ diagnostic.

To characterize the nature of cosmic expansion (whether it is accelerating or decelerating), we introduce the deceleration parameter $q$, defined as
\begin{equation} \label{q}
q = \frac{d}{dt}\left(\frac{1}{H}\right) - 1 = -\frac{\dot{H}}{H^2} - 1.    
\end{equation}

By substituting Eq. (\ref{Hz}) into Eq. (\ref{q}), the deceleration parameter takes the form
\begin{equation}\label{qz}
q(z)=-1+\frac{A\big(A H_0 + B\big)(1+z)^{-A}}
{B - \big(A H_0 + B\big)(1+z)^{-A}}. 
\end{equation}

In addition, the transition from decelerated to accelerated expansion is characterized by the condition $q(z_t)=0$. From Eq. \eqref{qz}, we have
\begin{equation}
\label{zt}
z_t=\left[\frac{(A+1)(A H_0+B)}{B}\right]^{1/A}-1.
\end{equation}

\begin{figure}[H]
    \centering
    \includegraphics[width=1.1\linewidth]{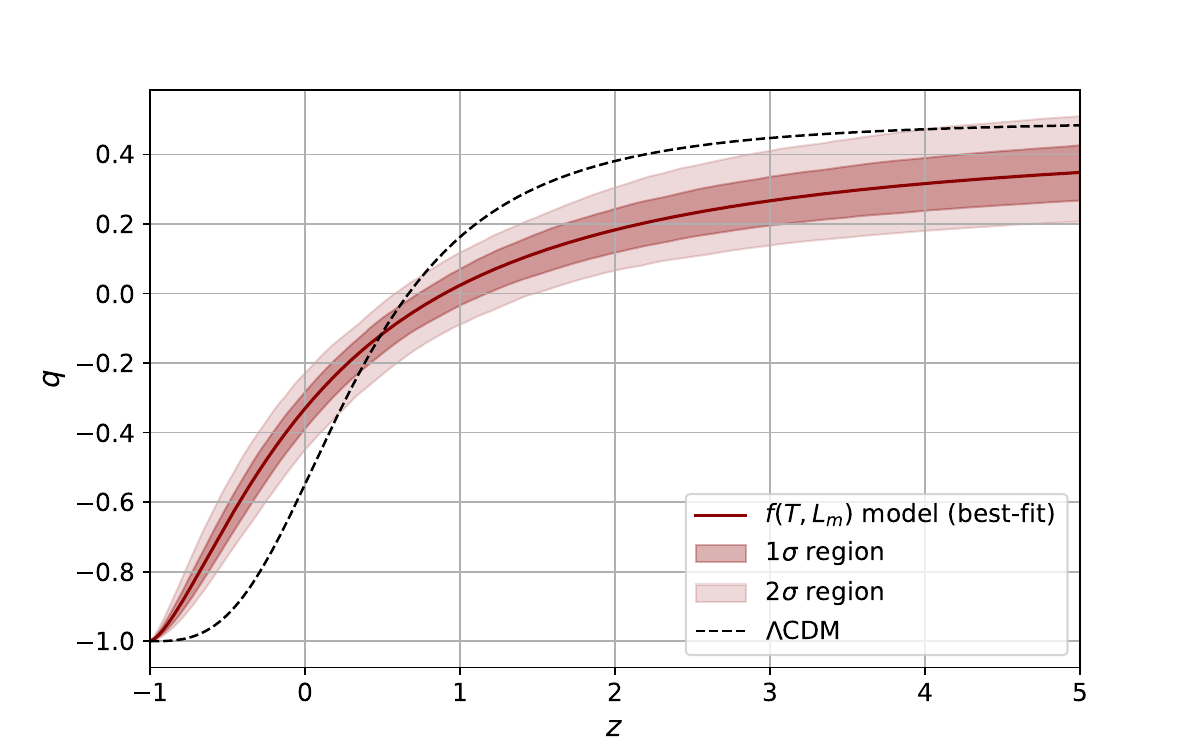}
    \caption{Evolution of the deceleration parameter $q(z)$ for the viscous $f(T, L_m)$ model compared with $\Lambda$CDM using the combined $H(z)$+SNe+DESI datasets.}
    \label{F_q}
\end{figure}

The value and sign of the deceleration parameter $q$ characterize the universe’s expansion dynamics: $q > 0$ indicates deceleration, reflecting a slowing expansion; $q < 0$ corresponds to acceleration; and $q = 0$ describes uniform expansion at a constant rate. The special case $q = -1$ represents the exact de Sitter expansion, featuring a constant Hubble parameter and purely exponential growth. Values $q < -1$ denote super-exponential expansion, surpassing de Sitter behavior. In Fig. \ref{F_q}, we depict the redshift evolution of the deceleration parameter $q(z)$ for the viscous $f(T,L_m)$ model, obtained using the best-fit parameters from the joint $H(z)$+SNe+DESI analysis. The solid red curve represents the model’s prediction, with the shaded regions corresponding to the $1\sigma$ and $2\sigma$ confidence intervals, while the dashed black line shows the evolution predicted by the standard $\Lambda$CDM model. At present time $(z=0)$, the deceleration parameter takes the value $q_0 = -0.33 \pm 0.41$, indicating an accelerating cosmic expansion consistent with observational evidence \cite{Campo_2012,Mamon_2018,Mamon_2016,Xu_2009}. As redshift increases, $q(z)$ transitions from negative to positive values, signifying a shift from the current accelerated phase to a past decelerating matter-dominated era. Further, we find that the transition redshift in the viscous $f(T,L_m)$ model, defined by $q(z_t)=0$, occurs at $z_t = 0.88^{+2.10}_{-0.78}$. The relatively large uncertainty reflects the weak constraints on the torsion-matter coupling parameter $\beta$, which enters nonlinearly into the expression for the deceleration parameter. In the far future $(z \to -1)$, the model predicts that $q(z) \to -1$, suggesting that the Universe asymptotically approaches a de Sitter phase dominated by dark energy. Hence, the behavior of $q(z)$ confirms that the viscous $f(T,L_m)$ model successfully describes the observed expansion history, remaining compatible with $\Lambda$CDM within the $2\sigma$ confidence level while introducing mild deviations due to modified gravity effects.

Further, to describe the dominant component of the Universe and its impact on cosmic expansion, we introduce the effective EoS, which relates the pressure $p_v$ of a viscous cosmic fluid to its energy density $\rho$ via:
\begin{equation} \label{w}
\omega_v=\frac{p_v}{\rho} = - \frac{\zeta_0 (\beta + 16 \pi)}{2 \alpha H(z)}.    
\end{equation}

By inserting Eq. (\ref{Hz}) into Eq. (\ref{w}), we get the effective equation of state parameter as
\begin{equation}
\omega_v(z) = - \frac{\zeta_0 (\beta + 16 \pi)}{2 \alpha \left[ -\frac{B}{A} + \left(H_0 + \frac{B}{A}\right) (1+z)^{-A} \right]}.    
\end{equation}

\begin{figure}[H]
    \centering
    \includegraphics[width=1\linewidth]{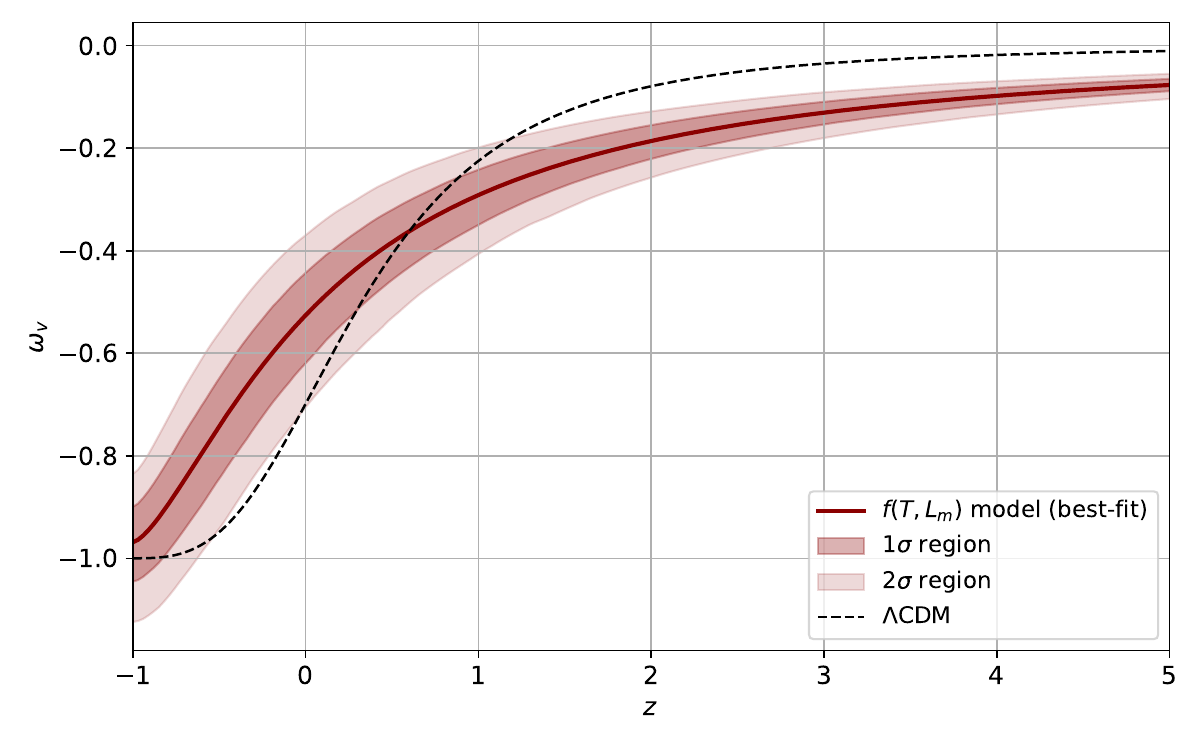}
    \caption{Evolution of the effective viscous EoS parameter $\omega_v(z)$ for the viscous $f(T, L_m)$ model compared with $\Lambda$CDM using the combined $H(z)$+SNe+DESI datasets.}
    \label{F_w}
\end{figure}

The value of $\omega_m = 0$ corresponds to pressureless matter (dust), typical of cold dark matter; $\omega_r = 1/3$ represents radiation, including photons and relativistic particles; $\omega_{\Lambda} = -1$ describes the cosmological constant ($\Lambda$), driving exponential (de Sitter) expansion; $\omega_p < -1$ indicates phantom energy, leading to super-accelerated expansion; and $-1 < \omega_q < -1/3$ corresponds to quintessence-like dark energy, producing accelerated expansion. Using the joint $H(z)$+SNe+DESI constraints, Fig. \ref{F_w} shows the redshift evolution of the effective viscous EoS $\omega_v(z)$ for the best-fit viscous $f(T,L_m)$ model (solid red line) with 1$\sigma$ and 2$\sigma$ credible regions (shaded bands), and the $\Lambda$CDM expectation as a dashed line. The model predicts that $\omega_v$ approaches $-1$ in the far future $(z \to -1)$ and increases monotonically with redshift, taking less negative values at earlier epochs. This behavior indicates that the effective fluid acts as a quintessence-like component ($\omega_v > -1$) at present and gradually evolves toward a pressureless state ($\omega_v \to 0$) at higher redshifts, consistent with the transition to the matter-dominated era. At the present epoch, we obtain $\omega_{v,0} = -0.52 \pm 0.26$ \cite{Hernandez,Zhang}. The viscous contribution is responsible for the model’s small departure from the $\Lambda$CDM curve at low redshift, while the shaded bands show that, given current data, these deviations remain statistically modest (compatible at the $\sim2\sigma$ level). Hence, the plot indicates that viscous effects in the $f(T,L_m)$ framework can drive late-time acceleration with an evolving EoS that is broadly consistent with present observations.

The $Om(z)$ diagnostic provides a straightforward method to distinguish different dark energy models in cosmology. It is particularly simple because it depends only on the first derivative of the cosmic scale factor. In a spatially flat universe, the $Om(z)$ parameter is defined as \cite{Sahni/2008}
\begin{equation}
Om(z) = \frac{\left[\frac{H(z)}{H_0}\right]^2 - 1}{(1+z)^3 - 1}.
\end{equation}

\begin{figure}[H]
    \centering
    \includegraphics[width=1\linewidth]{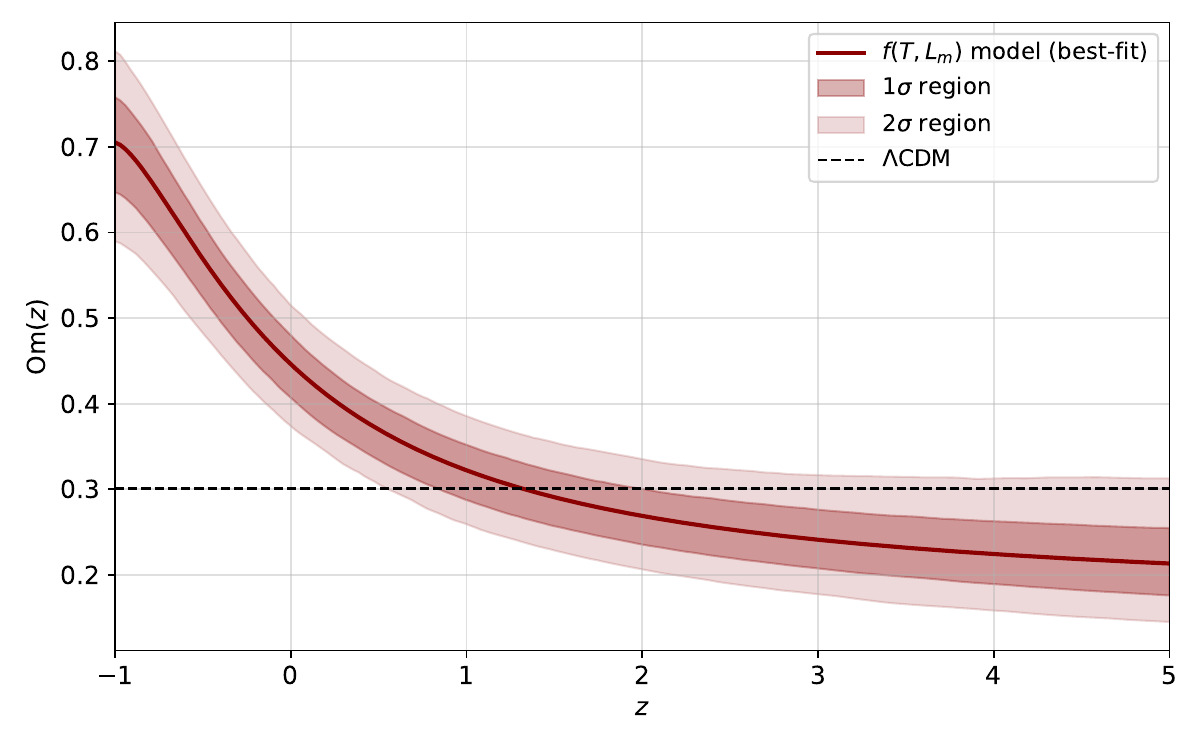}
    \caption{Evolution of the $Om(z)$ diagnostic for the viscous $f(T, L_m)$ model compared with $\Lambda$CDM using the combined $H(z)$+SNe+DESI datasets.}
    \label{F_Om}
\end{figure}

The negative slope of $Om(z)$ indicates quintessence-like behavior, whereas a positive slope signals phantom-like behavior. A constant value of $Om(z)$ corresponds to the standard $\Lambda$CDM model. Fig. \ref{F_Om} displays the evolution of the $Om(z)$ diagnostic for the viscous $f(T,L_m)$ model constrained by the combined datasets $H(z)$+SNe+DESI, compared with the standard $\Lambda$CDM prediction (black dashed line). In the $\Lambda$CDM framework, $Om(z)$ remains constant, equal to the present matter density parameter $\Omega_{m0}$. However, in the viscous $f(T,L_m)$ model (solid red curve), $Om(z)$ exhibits a clear redshift dependence, decreasing monotonically with $z$. The negative slope observed across the entire redshift range indicates a quintessence-type behavior ($\omega > -1$), meaning that the dark energy density gradually increases with cosmic time. At present ($z = 0$), the model predicts $Om(0) \approx 0.45$, which is somewhat higher than the $\Lambda$CDM value $\Omega_{m}^0 = 0.301$. This indicates that, while the model is broadly consistent with current observations, it allows for a mild deviation toward a dynamical dark energy behavior.

\section{Conclusion} \label{sec6}

In this study, we have investigated a cosmological scenario within the framework of viscous $f(T,L_m)$ gravity \cite{Harko2014a}, where the Universe is filled with a bulk viscous fluid instead of an ideal perfect fluid. The inclusion of bulk viscosity introduces an effective negative pressure term that can account for the observed late-time acceleration without invoking a cosmological constant. We adopted a simple linear model $f(T,L_m) = \alpha T + \beta L_m$, characterized by two coupling constants that quantify the torsion and matter–torsion interaction, and assumed a constant bulk viscosity coefficient $\zeta = \zeta_0 > 0$. Using recent observational datasets, including 31 Hubble parameter measurements from cosmic chronometers, the Pantheon+ sample of 1701 SNe Ia, and the latest DESI BAO data, we constrained the model parameters through a joint MCMC analysis. The best-fit values $H_0 = 68.16 \pm 0.65$, $\alpha = 1.53^{+0.49}_{-0.61}$, $\beta = 0.40 \pm 0.96$, and $\zeta_0 = 2.15^{+0.69}_{-0.81}$ are consistent with current cosmological observations. These results indicate a mild deviation from the standard teleparallel case and confirm that viscosity contributes significantly to the Universe’s accelerated expansion. The obtained $H_0$ value is in good agreement with Planck data, suggesting that the viscous $f(T,L_m)$ model provides a viable alternative explanation for dark energy within observational limits.

The reconstructed cosmological parameters show that the model reproduces the observed expansion history with high accuracy. The deceleration parameter $q(z)$ in Fig. \ref{F_q} evolves from positive to negative values, indicating a successful transition from a past decelerated phase to the current accelerated epoch, with $q_0 = -0.33 \pm 0.41$. In the far future ($z \to -1$), the model asymptotically approaches a de Sitter phase ($q \to -1$), confirming a sustained acceleration. The effective viscous EoS parameter $\omega_v(z)$ in Fig. \ref{F_w} evolves dynamically, taking the present value $\omega_{v,0} = -0.52 \pm 0.26$, consistent with a quintessence-like regime, and approaching $\omega_v \to -1$ in the future. Moreover, the $Om(z)$ diagnostic in Fig. \ref{F_Om} exhibits a decreasing trend, further supporting the quintessence-type behavior of the dark energy component. From a statistical perspective, the information criteria yield $\Delta \mathrm{AIC}=2.2$ and $\Delta \mathrm{BIC}=13.13$ relative to the $\Lambda$CDM model, indicating that although the viscous $f(T,L_m)$ model achieves a comparable goodness of fit to current observational data, the simplicity of $\Lambda$CDM remains statistically preferred when penalizing model complexity. Nevertheless, the viscous $f(T,L_m)$ framework offers a physically distinct and well-motivated description of late-time cosmic acceleration, driven by the combined effects of torsion-matter coupling and irreversible viscous processes, without the need for an explicit cosmological constant.

In summary, the viscous $f(T,L_m)$ framework represents a viable alternative to the standard cosmological model. It reproduces the observed late-time acceleration, remains consistent with current observational bounds, and naturally connects the effects of torsion, matter coupling, and dissipative processes within a unified theoretical context.

\section*{Acknowledgment}
This work was supported and funded by the Deanship of Scientific Research at Imam Mohammad Ibn Saud Islamic University (IMSIU) (grant number IMSIU-DDRSP2602).

\section*{Data availability}
The data sets used and/or analyzed during the current study are publicly available from the corresponding survey collaborations. No new data were generated in this work.


\begin{thebibliography}{} 

\bibitem{Perlmutter/1999} S. Perlmutter et al., \textit{Astrophys. J.} \textbf{517}, 565 (1999).

\bibitem{Riess/1998} A.G. Riess et al., \textit{Astron. J.} \textbf{116}, 1009 (1998).  

\bibitem{Riess/2004} A.G. Riess et al., \textit{Astrophys. J.} \textbf{607}, 665 (2004).  

\bibitem{Spergel/2003} D.N. Spergel et al., \textit{Astrophys. J. Suppl.} \textbf{148}, 175 (2003).  

\bibitem{D.J.} D.J. Eisenstein et al.. \textit{Astrophys. J.} \textbf{633}, 560 (2005).

\bibitem{W.J.} W.J. Percival et al.. \textit{Mon. Not. R. Astron. Soc.} \textbf{401}, 2148 (2010).

\bibitem{Koivisto/2006} T. Koivisto, D.F. Mota, \textit{Phys. Rev. D} \textbf{73}, 083502 (2006).  

\bibitem{Daniel/2008} S.F. Daniel et al., \textit{Phys. Rev. D} \textbf{77}, 103513 (2008).  

\bibitem{Peebles_2003} P. J. E. Peebles and B. Ratra, {\it Rev. Mod. Phys.} {\bf 75}, 559 (2003).  
\bibitem{Padmanabhan_2003} T. Padmanabhan, {\it Phys. Rep.} {\bf 380}, 235 (2003). 

\bibitem{Zlatev/1999} I. Zlatev, L. Wang, and P.J. Steinhardt, \textit{Phys. Rev. Lett.} \textbf{82}, 896 (1999).

\bibitem{Weinberg/1989} S.Weinberg,  \textit{Rev. Mod. Phys.} \textbf{61}, 1 (1989).

\bibitem{Buchdahl_1970} H. A. Buchdahl, {\it Mon. Not. R. Astron. Soc.} {\bf 150}, 1 (1970).  

\bibitem{Barrow_1983} J. D. Barrow and A. C. Ottewill, {\it J. Phys. A: Math. Gen.} {\bf 16}, 2757 (1983).  

\bibitem{Boehmer_2008} C. G. Boehmer, T. Harko, and F. S. N. Lobo, {\it Astropart. Phys.} {\bf 29}, 386 (2008). 

\bibitem{Bertolami_2007} O. Bertolami, C. G. Boehmer, T. Harko, and F. S. N. Lobo, {\it Phys. Rev. D} {\bf 75}, 104016 (2007).  

\bibitem{Harko_2008} T. Harko, {\it Phys. Lett. B} {\bf 669}, 376 (2008).

\bibitem{Harko_2010} T. Harko and F. S. N. Lobo, {\it Eur. Phys. J. C} {\bf 70}, 373 (2010).

\bibitem{Harko_2011} T. Harko, T. S. Koivisto, and F. S. N. Lobo, {\it Mod. Phys. Lett. A} {\bf 26}, 1467 (2011).  

\bibitem{Harko_2011_fRT} T. Harko, F. S. N. Lobo, S. Nojiri, and S. D. Odintsov, {\it Phys. Rev. D} {\bf 84}, 024020 (2011). 

\bibitem{Weitzenbock_1923} R. Weitzenb\"ock, {\it Invariantentheorie} (Noordhoff, Groningen, 1923). 

\bibitem{Einstein_1928} A. Einstein, {\it Sitzungsberichte der Preussischen Akademie der Wissenschaften, Phys.-math. Klasse}, 217 (1928).  

\bibitem{Moller_1961} C. Möller, {\it Mat. Fys. Skr. Dan. Vid. Selsk.} {\bf 1}, 10 (1961).  

\bibitem{Pellegrini_1963} C. Pellegrini and J. Plebanski, {\it Mat. Fys. Skr. Dan. Vid. Selsk.} {\bf 2}, 4 (1963).  

\bibitem{Hayashi_1979} K. Hayashi and T. Shirafuji, {\it Phys. Rev. D} {\bf 19}, 3524 (1979).  

\bibitem{Aldrovandi_2013} R. Aldrovandi and J. G. Pereira, {\it Teleparallel Gravity}, Fundamental Theories of Physics, Vol. 173 (Springer, Heidelberg, 2013).  

\bibitem{Ferraro2007} R. Ferraro and F. Fiorini, {\it Phys. Rev. D} {\bf 75}, 084031 (2007).

\bibitem{Ferraro2008} R. Ferraro and F. Fiorini, {\it Phys. Rev. D} {\bf 78}, 124019 (2008).

\bibitem{Bengochea2009} G. R. Bengochea and R. Ferraro, {\it Phys. Rev. D} {\bf 79}, 124019 (2009).

\bibitem{Linder2010} E. V. Linder, {\it Phys. Rev. D} {\bf 81}, 127301 (2010).

\bibitem{Boehmer2012} C. G. Böhmer, T. Harko, and F. S. N. Lobo, {\it Phys. Rev. D} {\bf 85}, 044033 (2012).

\bibitem{Harko2014b} T. Harko, F. S. N. Lobo, G. Otalora, and E. N. Saridakis, {\it JCAP} {\bf 12}, 021 (2014).

\bibitem{Bahamonde2017} S. Bahamonde, C. G. Böhmer, and M. Krššák, {\it Phys. Lett. B} {\bf 775}, 37 (2017).

\bibitem{Capozziello2017} S. Capozziello, G. Lambiase, and E. N. Saridakis, {\it Eur. Phys. J. C} {\bf 77}, 576 (2017).

\bibitem{Combi2018} L. Combi and G. E. Romero, {\it Ann. Phys.} {\bf 530}, 1700175 (2018).

\bibitem{Farrugia2018} G. Farrugia, J. L. Said, V. Gakis, and E. N. Saridakis, {\it Phys. Rev. D} {\bf 97}, 124064 (2018).

\bibitem{Awad2018} A. Awad, W. El Hanafy, G. G. L. Nashed, S. D. Odintsov, and V. K. Oikonomou, {\it JCAP} {\bf 07}, 026 (2018).

\bibitem{Myrzakulov2025} K. Myrzakulov, O. Donmez, M. Koussour, S. Muminov, E. Davletov, and J. Rayimbaev, {\it Phys. Lett. A} {\bf 534}, 130232 (2025).

\bibitem{Khokhar2025} U. A. Khokhar and Z. Yousaf, {\it Eur. Phys. J. Plus} {\bf 140}, 325 (2025).

\bibitem{Bajardi2025} F. Bajardi, D. Blixt, and S. Capozziello, {\it Phys. Rev. D} {\bf 111}, 084012 (2025).

\bibitem{Harko2014a} T. Harko, F. S. N. Lobo, G. Otalora, and E. N. Saridakis, {\it Phys. Rev. D} {\bf 89}, 124036 (2014).

\bibitem{Eckart/1940} C. Eckart, \textit{Phys. Rev.} \textbf{58}, 919-924 (1940).

\bibitem{Muller/1967} I. Müller, \textit{Zeitschrift für Physik} \textbf{198}, 329-344 (1967).

\bibitem{Israel/1976} W. Israel, \textit{Ann. Phys.} \textbf{100}, 310-331 (1976).

\bibitem{Israel/1979} W. Israel and J.M. Stewart, \textit{Ann. Phys.} \textbf{118}, 341-372 (1979).

\bibitem{Brevik/2005} I. Brevik, O. Gorbunova, and Y.A. Shaido, \textit{Int. J. Mod. Phys. D} \textbf{14}, 1899-1906 (2005).

\bibitem{Brevik/2006a} I. Brevik, \textit{Gen. Relativ. Gravit.} \textbf{38}, 1317-1328 (2006).

\bibitem{Brevik/2006b} I. Brevik, \textit{Int. J. Mod. Phys. D} \textbf{15}, 767-775 (2006).

\bibitem{Brevik/2012} I. Brevik, \textit{Entropy} \textbf{14}, 2302-2310 (2012).

\bibitem{ryden/2003} B. Ryden, \textit{ Introduction to Cosmology} (Addison Wesley, San Francisco, United States of America, 2003).

\bibitem{Sasidharan/2015} A. Sasidharan and T. K. Mathew, \textit{Eur. Phys. J. C} \textbf{75}, 348 (2015).

\bibitem{Ren/2006} J. Ren and X.-H. Meng, \textit{Phys. Lett. B} \textbf{633}, 1 (2006).

\bibitem{emcee} D. F. Mackey et al.,  Publ. Astron. Soc. Pac., \textbf{125}, 306 (2013).

\bibitem{Moresco/2012} M. Moresco et al., \textit{J. Cosmol. Astropart. Phys.} \textbf{08}, 006 (2012). 

\bibitem{Moresco/2015} M. Moresco, \textit{Mon. Not. R. Astron. Soc.} \textbf{450}, L16-L20 (2015).

\bibitem{Moresco/2016} M. Moresco, \textit{J. Cosmol. Astropart. Phys.} \textbf{05}, 014 (2016).

\bibitem{Moresco/2020} M. Moresco et al., \textit{Astrophys. J.} \textbf{898}, 82 (2020).

\bibitem{Scolnic_2022} D. M. Scolnic et al., \textit{Astrophys. J.} \textbf{938}, 113 (2022).

\bibitem{Brout_2022} D. Brout et al., {\it ApJ} {\bf 938}, 110 (2022).

\bibitem{BAO1} DESI collaboration, \textit{arXiv:2404.03000} (2024).

\bibitem{BAO2} A.G. Adame et al., \textit{J. Cosmol. Astropart. Phys.} \textbf{01}, 124 (2025).

\bibitem{BAO3} A.G. Adame et al., \textit{J. Cosmol. Astropart. Phys.} \textbf{02}, 021 (2025).

\bibitem{Planck_2015} P. A. Ade et al. (Planck Collaboration), {\it A\&A} {\bf 594}, A13 (2016).

\bibitem{Planck_2018} N. Aghanim et al. (Planck Collaboration), {\it A\&A} {\bf 641}, A6 (2020).

\bibitem{Riess_2016} A. G. Riess et al., {\it ApJ} {\bf 826}, 56 (2016).

\bibitem{Riess_2022} A. G. Riess et al., {\it ApJL} {\bf 934}, L7 (2022).

\bibitem{AIC1} H. Akaike, \textit{IEEE Trans. Autom. Control}  \textbf{19}, 716 (1974).

\bibitem{AIC2} G. Schwarz, \textit{Ann. Stat.} \textbf{6}, 461 (1978).

\bibitem{AIC3} D.J. Spiegelhalter, N. G. Best, B. P. Carlin, and A. van der Linde, \textit{J. R. Stat. Soc.} \textbf{64}, 583 (2002).

\bibitem{AIC4} K. Anderson, \textit{Model Selection and Multimodel Inference: A Practical Information-Theoretic Approach}, 2nd ed. (Springer, New York, 2002).

\bibitem{AIC5} K. P. Burnham and D. R. Anderson, \textit{Sociol. Methods Res.} \textbf{33}, 261 (2004).

\bibitem{AIC6} A. R. Liddle, \textit{Mon. Not. R. Astron. Soc.} \textbf{377}, L74 (2007).

\bibitem{AIC7} R. E. Kass and A. E. Raftery, \textit{J. Am. Stat. Assoc.} \textbf{90}, 773 (1995).

\bibitem{Campo_2012} S. del Campo et al., {\it Phys. Rev. D} {\bf 86}, 083509 (2012).

\bibitem{Mamon_2018} A. Al Mamon and K. Bamba, {\it Eur. Phys. J. C} {\bf 78}, 862 (2018).

\bibitem{Mamon_2016} A. Al Mamon and S. Das, {\it Int. J. Mod. Phys. D} {\bf 25}, 1650032 (2016).

\bibitem{Xu_2009} L. Xu, W. Li, and J. Lu, {\it J. Cosmol. Astropart. Phys.} {\bf 07}, 031 (2009), arXiv:0905.4552.

\bibitem{Hernandez} A. Hernandez-Almada, et al., \textit{Eur. Phys. J. C} \textbf{79}, 1-9 (2019).

\bibitem{Zhang} Q. J. Zhang, Y. L. Wu, \textit{J. Cosmol. Astropart. Phys.} \textbf{2010}, 038 (2010).

\bibitem{Sahni/2008}  V. Sahni, A. Shafieloo, and A.A. Starobinsky, \textit{Phys. Rev. D}, \textbf{78}, 103502 (2008).

\end{thebibliography}
\end{document}